\documentclass[journal,twoside,web]{ieeecolor}
\usepackage{generic}
\usepackage{cite}
\usepackage{amsmath,amssymb,amsfonts}
\usepackage{algorithmic}
\usepackage{graphicx}
\usepackage{algorithm,algorithmic}
\usepackage{hyperref}

\usepackage{multirow}
\usepackage{booktabs}
\usepackage{amssymb}
\usepackage{bbding}
\usepackage{pifont}
\usepackage{wasysym}
\usepackage{utfsym}
\usepackage{fontawesome}

\hypersetup{hidelinks=true}
\usepackage{textcomp}
\def\BibTeX{{\rm B\kern-.05em{\sc i\kern-.025em b}\kern-.08em
    T\kern-.1667em\lower.7ex\hbox{E}\kern-.125emX}}
\makeatletter
\newcommand{\@logo}{} \renewcommand{\@logo}{} 
\makeatother
\begin{document}
\title{ViG3D-UNet: Volumetric Vascular Connectivity-Aware Segmentation via 3D Vision Graph Representation}
\author{Bowen Liu, Chunlei Meng, Wei Lin, Hongda Zhang, Ziqing Zhou, Zhongxue Gan, Chun Ouyang$^{*}$
\thanks{This study was supported in part by the National Nature Science Foundation of China (No. 62201156), and in part by Shanghai Municipal Science and Technology Major Project (No. 2021SHZDZX0103). This work was also supported by the Shanghai Engineering Research Center of AI \& Robotics, Fudan University, China, and the Engineering Research Center of AI \& Robotics, Ministry of Education, China. The computations in this research were performed using the \textit{CFFF} platform of Fudan University. }
\thanks{All the authors are with Academy for Engineering and Technology, Fudan University, Shanghai 200433, China (e-mail:bwliu22@m.fudan.\protect\\
edu.cn; clmeng23@m.fudan.edu.cn; 22210860047@m.fudan.edu.cn;\protect\\ zhanghongda@fudan.edu.cn;    21110860021@m.fudan.edu.cn;    \protect\\ganzhongxue@fudan.edu.cn; oy\_c@fudan.edu.cn). }
}

\maketitle

\begin{abstract}
Accurate vascular segmentation is essential for coronary visualization and the diagnosis of coronary heart disease. This task involves the extraction of sparse tree-like vascular branches from the volumetric space. However, existing methods have faced significant challenges due to discontinuous vascular segmentation and missing endpoints. To address this issue, a 3D vision graph neural network framework, named ViG3D-UNet, was introduced. This method integrates 3D graph representation and aggregation within a U-shaped architecture to facilitate continuous vascular segmentation. The ViG3D module captures volumetric vascular connectivity and topology, while the convolutional module extracts fine vascular details. These two branches are combined through channel attention to form the encoder feature. Subsequently, a paperclip-shaped offset decoder minimizes redundant computations in the sparse feature space and restores the feature map size to match the original input dimensions. To evaluate the effectiveness of the proposed approach for continuous vascular segmentation, evaluations were performed on two public datasets, ASOCA and ImageCAS. The segmentation results show that the ViG3D-UNet surpassed competing methods in maintaining vascular segmentation connectivity while achieving high segmentation accuracy. Our code will be available soon. 
\end{abstract}


\section{Introduction}
\label{sec:introduction}
\IEEEPARstart{C}{ardiovascular} disease is recognized as a major global health concern \cite{roth2018global}. Among all cardiovascular diseases, coronary heart disease is the most prevalent and primarily results from abnormal coronary artery stenosis \cite{cooper2000trends}. 
Computed Tomography Angiography (CTA) is a high resolution non-invasive 3D imaging technique used to diagnose and plan the treatment of coronary artery disease \cite{serruys2023CAD}. Reconstruction of the vascular network from CTA images provides a crucial foundation for quantitative analysis of coronary stenosis and ensures a high confidence diagnosis. 
Before 3D reconstruction of blood vessels, accurate volumetric segmentation of the coronary arteries is essential to observe vascular spatial information \cite{serruys2023CAD}. 
Unlike abdominal organs, which are typically represented as clusters or masses, vessels such as coronary arteries and intracranial vessels appear as tree-like or tubular structures in volumetric space \cite{2022LiRuikunGraph-Connectivity}. This characteristic results in an overall sparse vessel that is dense and continuous along the vascular branches, making accurate vascular segmentation challenging.

In recent decades, considerable progress has been made in tubular structure detection methods, particularly in vessel segmentation. Learning-based automatic segmentation methods \cite{ronneberger2015unet,liu2021reviewMISeg} have demonstrated excellent performance in organ segmentation tasks, attributed to significant advances in deep learning within the semantic segmentation domain. 
With the evolution of Convolutional Neural Networks (CNN) from 2D to 3D, end-to-end 3D vascular segmentation \cite{isensee2021nnUnet} is now considered feasible. 
The introduction of attention mechanisms \cite{valanarasu2021GatedAxial-Attention}, followed by Vision Transformer (ViT) \cite{hatamizadeh2021SwinUnetr} modules, has enabled correlations between organs across different cross-sectional slices to be learned by models.
Additionally, research incorporating Graph Neural Networks (GNN) \cite{2022LiRuikunGraph-Connectivity,zhao2022GraphConvolution} indicates that segmentation performance is significantly improved by utilizing the geometric structure of the target object. 
Recently, the achievements of large language models in medical image segmentation \cite{SAM4MIS} have led to an enhancement in the ability of segmentation models to learn the morphology, connectivity, and positional priors of various organs in both 2D and 3D segmentation tasks.

\begin{figure*}[htb]
\centering
\includegraphics[width=\linewidth]
{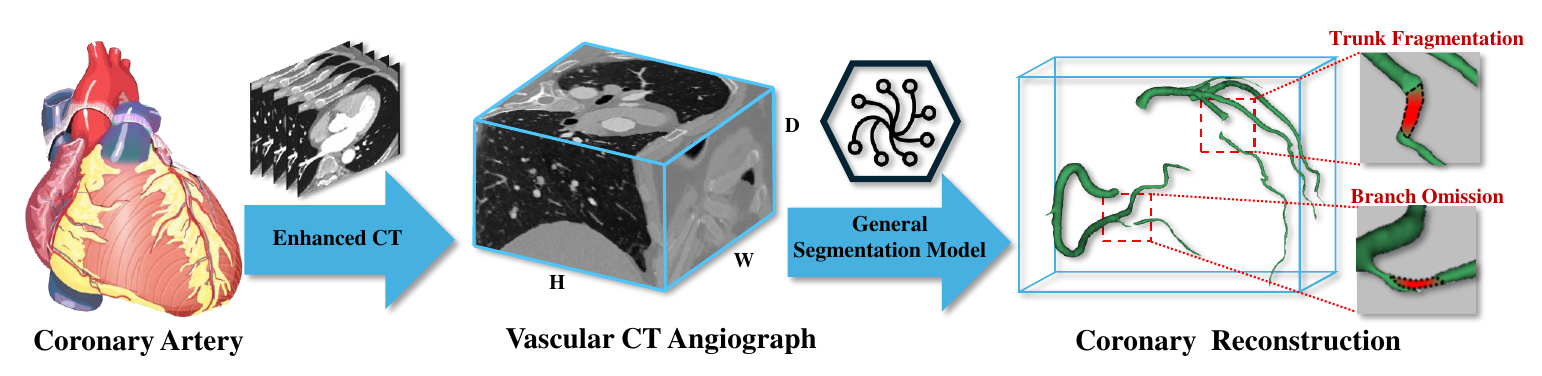}
\caption{Limitations of general-purpose coronary CTA segmentation: vascular discontinuity and incomplete distal branch detection.}
\label{fig0_intro}
\end{figure*}


Most existing general medical image segmentation models exhibit significant topological errors in vascular segmentation, as shown in Fig. \ref{fig0_intro}, characterized by fragmentation of primary vascular structures and omission of distal branches. 
These challenges primarily stem from three inherent characteristics of three-dimensional vascular images: (1) spatial sparsity of vascular networks, (2) diminished volumes of distal vessels \cite{2024TMI_EnergyMatchingVesselSeg} , and (3) ill-defined boundaries \cite{shamsi2024Topological}.
These limitations are exacerbated by conventional models' inability to simultaneously handle tubular geometry priors and microvascular contrast variations.
First, the U-Net framework and its variants struggle to capture all low-level features within topologies and fuzzy boundaries \cite{wang2022uctransnet}, which often leads to breaks in the segmentation results of slender branches and complex morphologies.
Second, volumetric segmentation usually employs patches as input due to the high computational demands of deep learning networks \cite{hatamizadeh2022unetr,hatamizadeh2021SwinUnetr} . This patch-based approach frequently neglects the integration of prior knowledge about vascular topology from the trunk to the branches.
Moreover, compared to CT medical images of abdominal and chest organs, the availability of 3D tubular structure datasets, particularly for coronary CTA data, is limited. Insufficient coronary data makes it inapplicable to train or fine-tune a large language model for 3D vessel segmentation tasks. 
Consequently, the development of an accurate extraction method that performs well in vascular structures is considered a significant challenge.

In this study, a \textbf{3D} \textbf{Vi}sion \textbf{G}raph based \textbf{U}-shaped neural \textbf{Net}work (ViG3D-UNet) framework was proposed for the continuous segmentation of coronary arteries. 
The encoder integrates a vision GNN to extract topological structures and a 3D CNN to capture textural features, respectively. These features, refined and fused through a channel attention module, are further processed by an offset decoder, designed to enhance the spatial connectivity of  the vascular segmentation results.
Initially, the \textbf{3D} \textbf{Vi}sion \textbf{G}raph neural network (ViG3D) module is crafted to establish graph representations, enabling the extraction of spatial vascular connectivity and topology. 
Subsequently, a 3D convolutional module is employed to capture vascular contextual features, functioning as a parallelized encoder alongside the ViG3D module. 
Finally, a offset decoder uniquely shaped like a paperclip, is utilized to accurately reconstruct high-dimensional vascular structures and texture features in the segmentation outcomes. The approach is validated through experiments on two publicly available coronary Computed Tomography Angiography (CTA) datasets, assessing the method's completeness and consistency in vascular segmentation tasks.
In summary, our contributions are as follows:
\begin{itemize}
    \item A 3D graph representation module, ViG3D, is proposed for volumetric segmentation, effectively ensuring connectivity and continuity in vascular segmentation results.
    \item A paperclip-shaped offset decoder is proposed to integrate ViG3D and 3D CNN features, reducing computational redundancy associated with 3D operations and improving the synergy between these network types.
    \item  Evaluations of two publicly available datasets demonstrate a substantial improvement in segmentation connectivity while preserving accuracy.
\end{itemize}

\section{Related Work}
\label{sec:guidelines}

\subsection{3D Networks for Segmentation}
With the development of U-shaped skip connection networks \cite{ronneberger2015unet}, CNN-based symmetric encoder-decoder structures \cite{cciccek20163dunet} have become the mainstream method for medical image segmentation. In CNN-based networks, deeper architectures \cite{huang2023Stu-net} and denser connections between layers \cite{zhou2019unet++} are employed to improve the performance of segmentation algorithms. The attention mechanism \cite{zhang2020ACSNet} was introduced for more effective aggregation of contextual information. Although CNN-based networks have improved segmentation accuracy \cite{isensee2021nnUnet,ExudateSeg}, CNN modules still do not extract sufficient contextual information adequately. A single CNN model is deemed inadequate to address the spatial sparsity and topological continuity inherent in vascular segmentation tasks.
Driven by transformers, the Vision Transformer (ViT) was developed by Dosovitskiy \textit{ et al.} \cite{2021ViT} for image tasks. Transformer methods are noted for their excellence in segmentation due to their capability to manage long-range dependencies. 
Unetr, a transformer framework for 3D medical image segmentation, was first proposed by Hatamizadeh \textit{et al.} \cite{hatamizadeh2022unetr}. 
Gated axial attention was introduced into ViTs by Valanarasu \textit{et al.} \cite{valanarasu2021MedicalTransformer} to improve computational efficiency.
The Swin Transformer, which employs a shifted window mechanism to reduce computational complexity in 3D segmentation, was proposed by Hatamizadeh \textit{et al.} \cite{hatamizadeh2021SwinUnetr}.
However, transformer network training requires large datasets, and vascular segmentation datasets are notably scarce, which poses challenges for ViTs in volumetric vascular segmentation.

\subsection{Tubular Structures Graph Representation}
Vessels typically exhibit a tubular structure. Various methods have been proposed by researchers \cite{wolterink2019coronaryCenterline,jiang2024ori-net,wang2020DDT-Net} to leverage this prior information, including the geometry and characteristics of tubular structures.
Wolterink \textit{et al.} \cite{wolterink2019coronaryCenterline} introduced a coronary artery tracker by following the centerline of the vessel. 
Jiang \textit{et al.} \cite{jiang2024ori-net} modeled coronary arteries as cylindrical structures and proposed an orientation-guided tracking method to enhance connectivity in coronary artery segmentation from CTA images. 
Wang \textit{et al.} \cite{wang2020DDT-Net} noted that tubular structures can be seen as a series of spheres and suggested a geometry-aware deep-distance transform network for tubular structure segmentation. 
Mou \textit{et al.} \cite{mou2021cs2net} developed CS2-Net, a segmentation network that incorporates self-attention mechanisms in curvilinear structures. 
Pan \textit{et al.} \cite{2024TMI_EnergyMatchingVesselSeg} proposed an energy-matching segmentation framework with an energy-based loss function to capture long-term vascular geometric information. Using tubular prior information improves segmentation accuracy, however, it is hard to ensure the continuity in segmentation tasks.

Tubular vessels can be further abstracted into graph data, with the vessels representing the edges and the branching points representing the vertices.
Han \textit{et al.} \cite{han2022ViG} introduced graph convolution for visual tasks and proposed vision GNN. 
Meng \textit{et al.} \cite{meng2021GraphRegionAggregation} built a GNN-based deep learning framework with multiple graph reasoning modules to leverage region and boundary features.
However, traditional GNNs are inherently suitable for planar data and cannot be directly applied to spatial data.
Antonio \textit{et al.} \cite{garcia2019UNet-graph3D} designed a UNet-GNN architecture for airway segmentation by replacing the deepest convolutional layers with graph convolutions.
Li \textit{et al.} \cite{2022LiRuikunGraph-Connectivity} proposed a graph attention network to model the graphical connectivity information of hepatic vessels.
Zhao \textit{et al.} \cite{zhao2022ShenRui} proposed a two-stage cascade segmentation model. In the first stage, a CNN-based approach is used to establish the vascular graph, while the second stage employs graph convolution to refine the results of coronary segmentation.
3D vascular images contain densely packed voxels, making it difficult to directly construct vascular graphs from raw 3D data. Graph network segmentation typically involves two steps: graph construction and segmentation. As a result, graph-based segmentation frameworks are inherently two-stage, further increasing the complexity of training and application.

\subsection{Feature Fusion for Vascular Segmentation}
In vascular segmentation tasks, the texture and topology of blood vessels are crucial characteristics. However, the relationship between these features is often overlooked by deep models when trained individually. 
Zhang \textit{et al.} \cite{2018ExFuse} introduced semantic information into low-level features and high-resolution details into high-level features for semantic segmentation.
Nagaraj \textit{et al.} \cite{zhang2022featureFusion} proposed a multi-path feature fusion for medical image segmentation.
Wu \textit{et al.} \cite{2021SCS-Net} designed a scale- and context- sensitive network to capture representative and distinguishing features of retinal vessels. 
Due to differences in scale and form of features between graph networks and CNNs, approaches for integrating dense 3D CNN features with sparse graph network features have not yet been developed.

\section{Methodology}

\begin{figure*}[hbtp]
\centering
\includegraphics[width=\linewidth]
{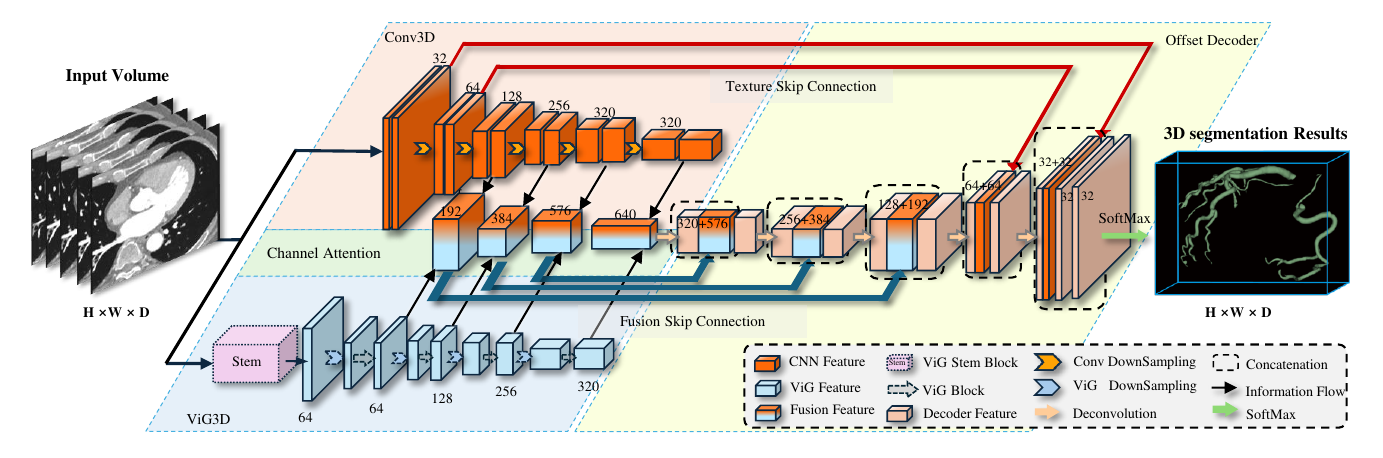}
\caption{Overview of ViG3D-UNet Architecture. The encoder is composed of a 3D vision GNN module and a 3D CNN module.The two modules are combined into a parallelized encoder under the block of channel attention. The fusion feature obtained via channel attention is concatenated with the first three layers of the decoder through skip connections to form the decoder feature. The first two layers of texture features, obtained through 3D convolution, are concatenated with the last two layers of the decoder through skip connections to form the decoder feature.}
\label{fig1 overview}
\end{figure*}

To address the spatial vascular segmentation task, which demands spatial continuity and topological coherence in target voxels, we proposed a 3D vision graph-based U-shaped framework, ViG3D-UNet.
The overall framework, illustrated in Fig. \ref{fig1 overview}, is designed in an encoder-decoder configuration similar to a U-shaped network. 
It consists of a parallelized dual-branch encoder, a channel attention module, and a uniquely paperclip-shaped offset decoder.
Unlike traditional two-stage serial cascade graph network approaches \cite{zhao2022ShenRui}, this framework employs a dual-branch parallelized encoder to simultaneously integrate vascular texture and topology information.
The two encoder branches incorporate a 3D CNN module \cite{isensee2021nnUnet} and a 3D Vision GNN module, respectively. 
The ViG3D module functions as a spatial network module to enable end-to-end training through graph aggregation of vessel connectivity feature. 
The features of vascular texture are derived from the CNN module, while the topological features are captured by the ViG3D module.
The channel attention module enriches the vascular features by integrating these two data types. 
For enhanced computational efficiency, a paperclip-shaped offset decoder is used during the up-sampling, integrating textural features with topological features. 
Further details on the methodology are provided below.

\subsection{3D Vision GNN Module}
\begin{figure*}[htp]
\centering
\includegraphics[width=\linewidth]
{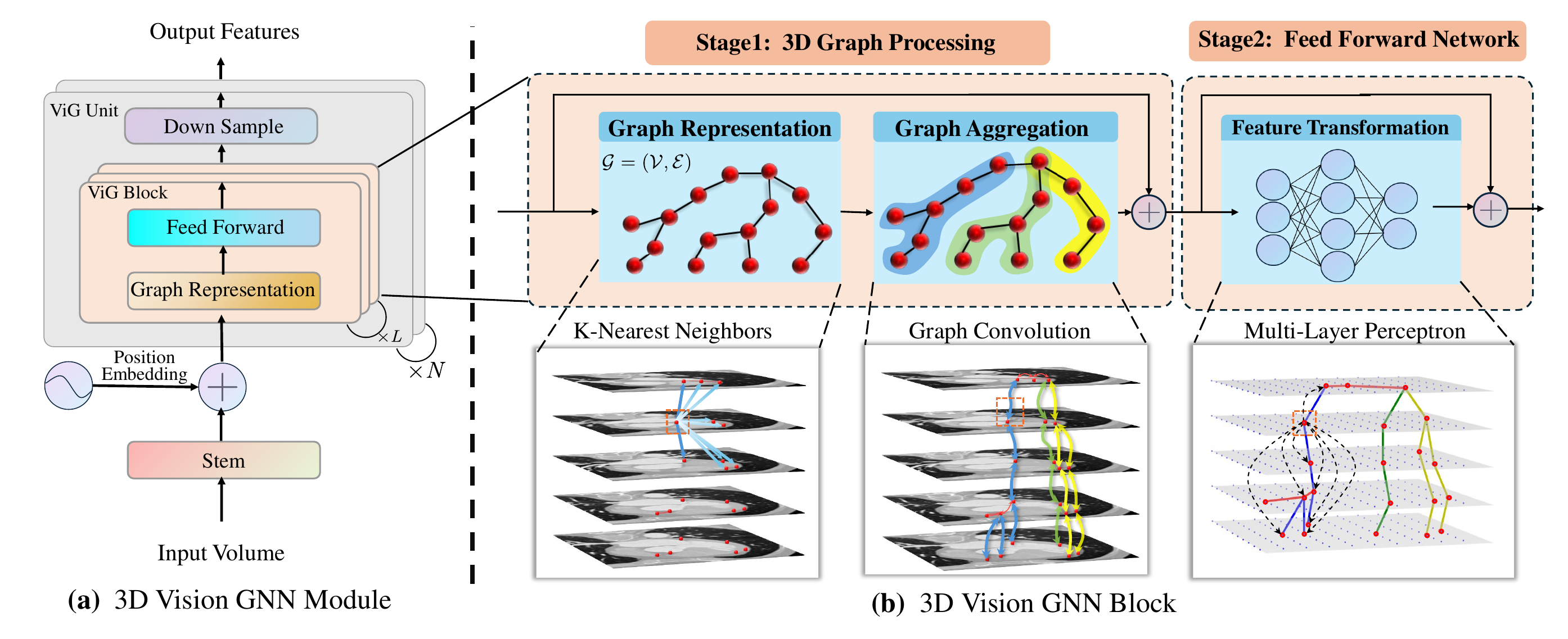}
\caption{Illustration of the 3D Vision GNN Module which identifies the connections between vascular nodes through graph aggregation operations during network training. (a) An overview of the branch of 3D Vision GNN in parallelized encoder. (b) Two stages in 3D Vision GNN Block: 3D graph processing and feed forward network.}
\label{fig2_3DVIG}
\end{figure*}

The 3D Vision GNN module, which is one of the encoder branches, is designed to extract the vascular connectivity features. 
 Initially, a stem block is used to reduce the dimensionality of the data and the vascular nodes of the sample, as illustrated in Fig. \ref{fig2_3DVIG}(a). 
 Positional embeddings are then added to these outputs, which are processed by the ViG3D module. Within this module, features undergo graph aggregation, transforming into graph features. 
 The graph representation, combined with the feed-forward operator, creates the core structure for the representation vascular graph  in the ViG3D block.
 Each ViG3D block is repeated $\mathit{L}$ times, followed by a down-sampling operator to form a single layer ViG3D unit. 
 The encoder branch on pyramid ViG3D module is constructed by repeating each ViG3D unit $\mathit{N}$ times on each feature scale. 

The stem block is composed of three convolutional layers. The first two layers, each with a stride of 2, effectively halve the input dimensions in terms of height, width, and depth. The final layer, with a stride of 1, further aggregates pixel-level features. 
As a result, the output features of the stem block are reduced to dimensions of $\frac{H}{4}$, $\frac{W}{4} $ and $\frac{D}{4}$, where $\mathit{H}$, $ \mathit{W}$ and $\mathit{D}$denote the original height, width, and depth of the input image.
This reduction in dimensionality decreases the computational load on subsequent 3D Vision GNN blocks and facilitates the conversion of dense voxels into a discrete form, optimizing them for a graph-based approach.

ViG3D block comprises two stages as depicted in Fig. \ref{fig2_3DVIG}(b): 3D graph processing and a forward network. In the graph processing stage, the input features are treated as a set of unordered vertices, represented by $\mathcal{V}=\left\{v_{1}, v_{2}, \cdots, v_{N}\right\}, i=1,2,\cdots,\mathit{N}$.The number of nodes per input layer is given by $\mathit{N}=\frac{H}{4}\times\frac{W}{4}\times\frac{D}{4}$. For each voxel $v_{i}$, its neighboring voxels $\mathcal{N}(v_j)$ were identified using the K-nearest neighbors method. Consequently, the edges $e_{ij}$ are formed from the node $v_{i}$ to each of its neighboring voxels $v_j \in \mathcal{N}\left ( v_j \right ) $.

The vascular nodes belonging to the same branch exhibit higher distance metrics. This graph representation enhances information transmission between adjacent vascular nodes during graph convolution operations. Once the graph representation is complete, the graph $\mathcal{G}=\left ( \mathcal{V},\mathcal{E}\right )$ of each input feature map is established. For the given input feature $\mathbf{X}\in \mathbb{R}^{N\times C} $, the graph representation process is subsequently referred to as $\mathcal{G}=G\left ( \mathbf{X}\right )$. The graph convolution operation $H_{GConv}$ can be expressed as:
\begin{align}
    \mathcal{G}'&=H_{GConv}\left ( \mathcal{G}(\mathit{X}), \mathit{W}\right )\nonumber \\
    &=H_{upd}\left ( H_{agg}(\mathcal{G}(\mathit{X}), \mathit{W_{agg}}), \mathit{W_{upd}}\right ),
\end{align}
where $H_{agg}$ and $H_{upd}$ denote the aggregation and update operations within the graph convolution process. The operators are parameterized by the learnable parameters $W_{agg}$ and $W_{upd}$\cite{huang2023vigu}. 

As the vascular graph is established, 3D graph aggregation is used to extract features of 3D tubular structures.
For each individual node $\mathbf{x}_{i}$, the 3D graph aggregation merges node features from  the surrounding spatial neighborhood.
The process of obtaining aggregated features $\mathbf{x}_{i}^{\prime}$ from input features $\mathbf{x}_{i}$ can be described as:
\begin{align}
    \mathbf{x}_{i}^{\prime}=h\left(\mathbf{x}_{i}, g\left(\mathbf{x}_{i}, \mathcal{N}\left(\mathbf{x}_{i}\right), W_{\text {agg}}\right), W_{\text {upd}}\right),
\end{align}
where $\mathcal{N}\left(\mathbf{x}_{i}\right)$ denotes the set of neighboring nodes of $\mathbf{x}_{i}$. The graph aggregation is specifically implemented using max-relative graph convolution $g\left( \cdot \right)$ due to its superior performance in vision tasks as demonstrated in \cite{li2019deepgcns}, expressed as:  
\begin{align}
    g\left( \cdot \right) =\left[\mathbf{x}_{i}, \max \left(\left\{\mathbf{x}_{j}-\mathbf{x}_{i} \mid j \in \mathcal{N}\left(\mathbf{x}_{i}\right)\right\}\right)\right].
\end{align}

For input feature $\mathbf{X}$, the graph processing stage can be denoted as:
\begin{align}
    \mathbf{X}^{\prime} = \text {GraphConv}\left( \mathbf{X}\right).
\end{align}
Edges that connect vascular nodes within the same branch enhance the representation of the feature, thus improving the continuity of the segmentation results for that vessel branch.

In the context of graph convolution, two multi-layer perceptrons are employed at both the input and output stages to project node features into a unified domain, thereby enhancing feature diversity. 
For input feature $\mathbf{X} $, the ViG3D block in graph processing stage can be expressed as:
\begin{align}
    \mathbf{Y}=\sigma (\text{GraphConv}(\mathbf{X}W_{in}))W_{out} + \mathbf{X},
\end{align}
where $\mathbf{Y} \in \mathbb{R}^{N\times C}$,$W_{in}$ and $W_{out}$ denote the input and output weights of the multi-layer perceptron, respectively. The activation function $\sigma$ is implemented using GeLU \cite{hendrycks2016GeLU}.

After the graph processing stage, the feature transformation is carried out by a feed-forward network. 
Each node is further processed by an additional multi-layer perceptron comprising two fully connected layers, which enhance connections between nodes within the same branch.
The 3D graph processing and feed-forward stages jointly constitute the ViG3D block, the core unit of the 3D Vision GNN encoder module.

\subsection{Parallelized Encoder}
The encoder comprises two parallelized pyramid branches: one that incorporates a CNN module and the other incorporates a ViG3D module. Table \ref{tab:encoder setting} details the settings of this dual-branch encoder, where $\mathit{C}$ is the dimension of the feature channel, $\mathit{S}$ is the convolution stride, $\mathit{E}$ is the number of convolutional layers in the feed-forward networks, and $\mathit{K}$ is the number of neighbors in the graph representation. 
In the CNN module, a 3D convolutional operator is used as the backbone to capture local texture information in coronary images.
This module includes six layers, designed to extract features at various resolutions: the 3D convolutional operators increase feature dimensions, while the down-sampling operators decrease feature size. 
In the ViG3D module,  the stem block comprises the convolutional layers found in $\text{Enc}\_1$ and $\text{Enc}\_2$, and the final layers consist of a stack of ViG3D blocks.

A channel attention module is used to merge vascular textural and topological information from the CNN and ViG branches.
As illustrated in Fig. \ref{fig1 overview}, feature maps of the CNN and ViG branches, specifically from $\text{Enc}\_3$ 
to $\text{Enc}\_6$,  are selected to integrate vascular connectivity and texture information for segmentation. 
First, the CNN branch features $F_{\mathit{CNN}}^{i} \in \mathbb{R}^{H\times W \times D\times C_{i}}$ are concatenated with the ViG branch features $F_{\mathit{ViG}}^{j} \in \mathbb{R}^{H\times W \times D\times C_{j}}$ to form merged features $F_{\mathit{M}} \in \mathbb{R}^{H\times W \times D\times (C_{i}+C_{j})}$. 
Next, these merged features undergo a series of transformations, including average pooling, a multi-layer perceptron, and a sigmoid activation function, culminating in the creation of a channel attention map $F_{\mathit{AM}} \in \mathbb{R}^{1\times 1 \times 1\times (C_{i}+C_{j})}$. 
This attention map is then multiplied by the combined features $F_{\mathit{AM}}$ in elemental order, effectively weighting each channel to highlight relevant vascular information while diminishing irrelevant details. 
The output of the channel attention module is $F_{\mathit{CA}} \in \mathbb{R}^{H\times W \times D\times (C_{i}+C_{j})}$. These features, enhanced through the channel attention mechanism, encompass richer semantic and topographical information, thereby improving their suitability for the decoder process.

\subsection{Offset Decoder}

Since most operators on the encoder are three-dimensional,
directly passing encoder features to the decoder via skip connections can introduce computational redundancy. 
The proposed decoder diverges from traditional U-shaped decoders by not relying solely on a single encoder for skip connections. 
Instead, as depicted in Fig. \ref{fig1 overview}, it integrates texture features from the 3D CNN module and fusion features from the channel attention module as primary inputs to the paperclip-shaped offset decoder.
Textural features are extracted from the first two layers of the encoder’s convolution branch and fed into the last two layers of the decoder, while fusion features, sourced from the last four layers of the ViG3D and 3D CNN modules, are directed to the first four layers of the decoder. 
Each layer of the decoder employs two convolutional layers to process the concatenated features, utilizing convolution, instance normalization \cite{ulyanov2016instanceNormal}, and ReLU activation. Each decoder block begins with a $3\times3\times3$ convolution, followed by a transposed convolution to achieve up-sampling.

\begin{table}[htbp]
\centering
\caption{Detailed Settings of Parallelized Encoder}
\label{tab:encoder setting}
\scalebox{0.92}{
\begin{tabular}{|c|c|c|c|}
\hline
Layer&Output Size&CNN Module&ViG3D Module\\

\hline
Enc\_1 & $\frac{H}{2}\times\frac{W}{2}\times\frac{D}{2}$ 
& $\begin{matrix} \text{Conv3d}\begin{bmatrix} C=32 \\ S=1 \end{bmatrix} \\ \text{Downsampling} \end{matrix}$ 
& $\text{Conv3d}\begin{bmatrix} C=32 \\ S=2 \end{bmatrix} \times1$\\

\hline
Enc\_2 & $\frac{H}{4}\times\frac{W}{4}\times\frac{D}{4}$ 
& $\begin{matrix} \text{Conv3d}\begin{bmatrix} C=64 \\ S=1 \end{bmatrix} \\ \text{Downsampling} \end{matrix}$ 
& $\begin{matrix} \text{Conv3d}\begin{bmatrix} C=64 \\ S=2 \end{bmatrix} \times1\\ \text{Conv3d}\begin{bmatrix} C=64 \\ S=1 \end{bmatrix} \times1 \end{matrix}$ \\

\hline
Enc\_3 & $\frac{H}{8}\times\frac{W}{8}\times\frac{D}{8}$ 
& $\begin{matrix} \text{Conv3d}\begin{bmatrix} C=128 \\ S=1 \end{bmatrix} \\ \text{Downsampling} \end{matrix}$ 
& $\begin{matrix} \text{ViG3D}\begin{bmatrix} C=64\\E=2\\K=9 \end{bmatrix} \times2 \\ \text{Downsampling}\end{matrix}$ \\

\hline
Enc\_4 & $\frac{H}{16}\times\frac{W}{16}\times\frac{D}{16}$ 
& $\begin{matrix} \text{Conv3d}\begin{bmatrix} C=256 \\ S=1 \end{bmatrix} \\ \text{Downsampling} \end{matrix}$ 
& $\begin{matrix} \text{ViG3D}\begin{bmatrix} C=128\\E=2\\K=9 \end{bmatrix} \times4 \\ \text{Downsampling}\end{matrix}$ \\

\hline
Enc\_5 & $\frac{H}{32}\times\frac{W}{32}\times\frac{D}{32}$ 
& $\begin{matrix} \text{Conv3d}\begin{bmatrix} C=320 \\ S=1 \end{bmatrix} \\ \text{Downsampling} \end{matrix}$ 
& $\begin{matrix} \text{ViG3D}\begin{bmatrix} C=256\\E=2\\K=9 \end{bmatrix} \times6 \\ \text{Downsampling}\end{matrix}$ \\

\hline
Enc\_6 & $\frac{H}{32}\times\frac{W}{32}\times\frac{D}{64}$ 
& $\begin{matrix} \text{Conv3d}\begin{bmatrix} C=320 \\ S=1 \end{bmatrix} \\ \text{Downsampling} \end{matrix}$ 
& $\begin{matrix} \text{ViG3D}\begin{bmatrix} C=320\\E=2\\K=9 \end{bmatrix} \times2 \\ \text{Downsampling}\end{matrix}$ \\
\hline
\end{tabular}
}
\end{table}

\section{Experimental Setup and Results}  
\subsection{Experimental Setup}
\subsubsection{Datasets and Description} 
The Automated Segmentation of Coronary Arteries (ASOCA) dataset, used in the MICCAI-2020 Challenge \cite{gharleghi2022ASOCA}, is a publicly accessible dataset tailored for coronary artery segmentation.  It consists of 40 annotated Cardiac CTA images with an initial resolution of $512\times512\times\mathit{N}$, where $\mathit{N}$ ranges from 168 to 334. This dataset is strategically divided into 30 samples for training and 10 samples for testing, focusing on the task of volumetric vascular segmentation.

The ImageCAS dataset, titled "A Large-Scale Dataset and Benchmark for Coronary Artery Segmentation" \cite{zeng2023imagecas}, is a comprehensive collection of 3D CTA images from 1000 patients diagnosed with coronary artery disease.
This dataset notably includes individuals who underwent early revascularization within 90 days after diagnosis, providing valuable information on post-treatment coronary anatomy. Each image is characterized by dimensions of $512\times512\times\mathit{N}$, where $\mathit{N}$ ranges from 206 to 275. The planar resolution varies between 0.29 and 0.43 mm$^{2}$, while the slice spacing varies from 0.25 to 0.45 mm. The dataset is strategically partitioned into 800 samples for training and 200 for testing, facilitating robust model validation.

\subsubsection{Implementation Details}
The optimal hyperparameters for the neighbor nodes in the ViG3D module were determined by primarily evaluating the performance of the method while adjusting the number of nodes for graph representation. Furthermore, the structure of ViG3D was verified through ablation studies, with specific details provided in the following sections.

The proposed model was trained on the ASOCA and ImageCAS datasets for 1,000 epochs, each comprising 250 iterations. We utilized the SGD optimizer with a polynomial decay scheduler, initiating the learning rate at $1e^{-2}$ and decaying at a rate of $3e^{-5}$ over 50,000 iterations. A five-fold cross-validation strategy was used for experiments on both datasets. A batch size of 2 was used, with the input patch dimensions set to $80\times192\times160$.
The network implementation was carried out using PyTorch and trained with the nnUNet framework \cite{isensee2021nnUnet}. All models were trained from scratch on two Tesla A100 40G GPU cards. The segmentation model used a combined loss function, comprising both Dice and cross-entropy components, defined as follows:
\begin{align}\mathcal{L}_{i}=\lambda_{1}\mathcal{L}_{Dice}+\lambda_{2}\mathcal{L}_{CE},  
\end{align}
where $\mathcal{L}_{i}$ represents the loss of decoder Layer $i$.  The weight coefficients $\lambda_{1}$ and $\lambda_{2}$ are set to 0.5. To alleviate the problem of the disappearance of the gradient, the deep supervision loss \cite{lee2015DS_loss} was applied to the last five decoders, with the loss weights successively halved from Decoder 6 to Decoder 2.

\subsubsection{Performance Metrics}
The Dice Similarity Coefficient (DSC) is employed to measures the accuracy of segmentation results by evaluating the overlap between predicted and actual segmentation maps, defined as:
\begin{align}
    DSC=\frac{2\times TP}{2\times TP+FP+FN}, 
\end{align}
where $TP$ (True Positive) represents the number of pixels correctly identified as part of the coronary class, $FP$ (False Positive) indicates pixels incorrectly identified as belonging to the coronary class, and $FN$ (False Negative) refers to pixels wrongly identified as the background class. The DSC metrics reflect how accurately the model captures coronary structures.

The Average Symmetric Surface Distance (ASSD) and the Hausdorff Distance (HD) 95\% were employed to assess boundary agreement between the segmentation results and the ground truth. The ASSD is calculated from the average surface distance, whereas the HD95 returns the maximum value within the 95th percentile of the Hausdorff distance measurements. The formulas for ASSD and HD are defined as follows:
\begin{align}
        ASSD=\frac{1}{\left | A \right |+\left | B \right |  }(\sum_{a\in A}^{}\min_{b\in B}(a,b)+\sum_{b\in B}^{}\min_{a\in A}(b,a)  ), 
\end{align}
\begin{align}
    HD=H(A,B)=max(h(A,B),\ h(B,A)),
\end{align}
where $A$ and  $B$ represent the sets of surface points for the predicted and ground truth segmentation, respectively. $\left | A \right |$ and $\left | B \right |$ represent the number of points in surfaces $A$ and $B$. $\min_{b\in B}(a,b)$ represents the minimum Euclidean distance from a point $a$ on the surface $A$ to the surface $B$, $h(A, B)$ is the maximum distance from any point in Surface $A$ to the nearest point in Surface $B$, and $h(B, A)$ is the maximum distance from any point in Surface $B$ to the nearest point in Surface $A$.

In addition, Intersection over Union (IoU), precision, and recall were selected as evaluation metrics.
The IoU is used as a supplementary metric to assess the accuracy of segmentation. Precision indicates the algorithm's ability to maintain connectivity in segmentation results, with a higher precision value suggesting better connectivity. Recall emphasizes the ability of the algorithm to detect all vascular regions, reflecting its coverage. A high recall rate indicates that the model successfully identified most of the vascular branches. These metrics are defined as follows:

\begin{align}
    IoU=\frac{TP}{TP+FP+FN}, 
\end{align}
\begin{align}
    Precision=\frac{TP}{TP+FP}, 
\end{align}
\begin{align}
    Recall=\frac{TP}{TP+FN}. 
\end{align}
 Better performance is indicated by higher DSC, IoU, precision and recall values, as well as lower ASSD and HD95.

\subsection{Segmentation Performance for Our Model}

Experiments of a five-fold cross-validation strategy on the ASOCA and ImageCAS datasets, respectively. The comprehensive evaluation metric is shown in Table \ref{tab:metrics_ours}. In the small-scale ASOCA dataset, the average values of DSC, ASSD, HD95, IoU, precision and recall values were 84.23\%, 1.85, 13.39, 73.06\%, 90.39\% and 79.65\%, respectively. The final column in Fig. \ref{fig3_ASOCA_results} shows visual segmentation examples of ViG3D-UNet for four cases.  In the large-scale ImageCAS dataset,  the average DSC, ASSD, HD95, IoU, precision and recall values were 81.72\%, 2.41, 15.21, 69.40\%, 83.16\% and 80.79\%, respectively. The final column in Fig. \ref{fig4_ImageCAS_results} shows visual segmentation examples of ViG3D-UNet for four cases. Both quantitative and qualitative results demonstrate that the proposed model achieves exceptional segmentation performance and maintains good vascular connectivity. Notably, on smaller-sample datasets, our method achieves better average metrics with a lower standard deviation. This reflects the significant segmentation performance of the proposed method when training data is limited.

\begin{table*}[!hbtp]
\centering
\caption{Coronary vessel segmentation performance metrics on ASOCA and ImageCAS datasets.} 
\label{tab:metrics_ours} 
\resizebox{\linewidth}{!}{
\begin{tabular}{l|cc|cc|cc|cc|cc|cc} 
\hline 
    &     \multicolumn{2}{c|}{\underline {\phantom{xxx}DSC ↑\phantom{xxx}}}&  \multicolumn{2}{c|}{\underline{\phantom{xxx}ASSD ↓\phantom{xxx}}}& \multicolumn{2}{c|}{\underline{\phantom{xxx}HD95 ↓\phantom{xxx}}}&  \multicolumn{2}{c|}{\underline{\phantom{xxxx}IoU ↑\phantom{xxxx}}}& \multicolumn{2}{c|}{\underline{\phantom{xx}Precision ↑\phantom{xx}}}&  \multicolumn{2}{c}{\underline{\phantom{xxx}Recall ↑\phantom{xxx}}}\\
 \rule{0pt}{2ex}    & ASOCA &ImageCAS  & ASOCA &ImageCAS  &ASOCA  &ImageCAS  & ASOCA &ImageCAS  & ASOCA &ImageCAS  & ASOCA &ImageCAS  \\   
    \hline
Mean& 84.23	&81.72	   &1.85  &2.41	  &13.39	  &15.21	     &73.07	 &69.40	    &90.39	&83.16	   &79.56  &80.79     \\
Std & 4.81	&5.09	   &1.91  &1.96	  &24.86	  &18.53	     &7.09	 &7.07	    &6.25	&6.04	   &8.34   &7.07      \\
Max & 90.62 &89.62	   &6.67  &13.26    &82.74	  &109.91     &82.85	 &81.20	    &96.30  &95.35     &94.89  &93.94     \\
\hline
\end{tabular}
}
\end{table*}

\subsection{Ablation Studies} 
\begin{figure}[!htbp]
\begin{center}
   \includegraphics[width=0.95\linewidth]{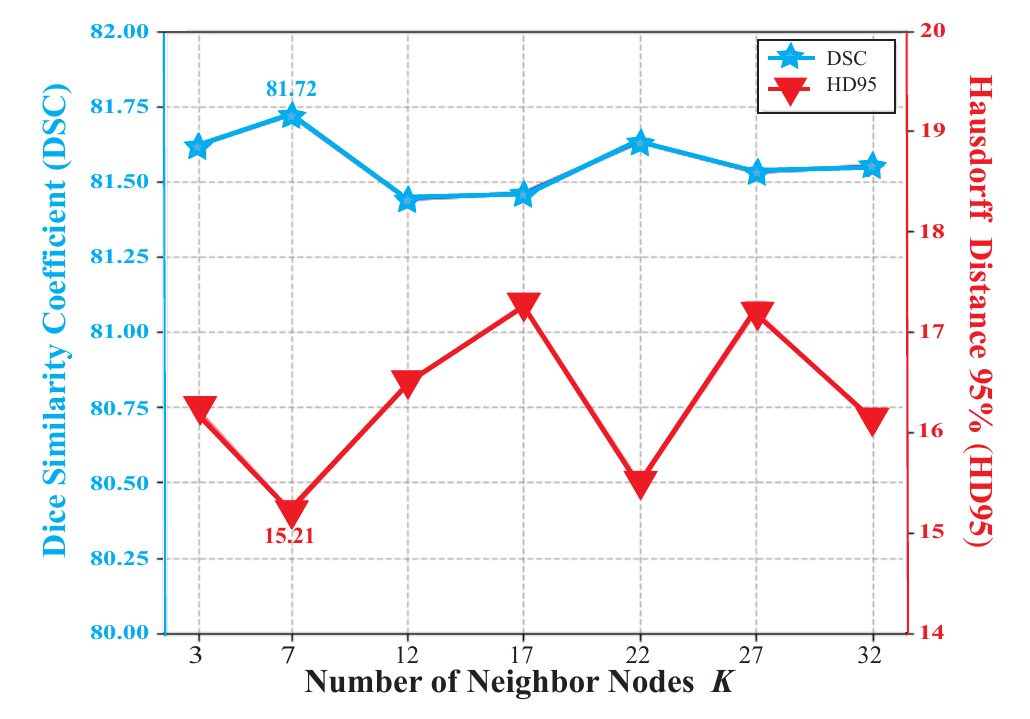}
\end{center}
   \caption{Performance comparison with varying numbers of neighbor nodes $\mathit{K}$ on the ImagesCAS dataset, with $\mathit{K}$ values ranging from 3 to 32. Optimal model performance is observed when $\mathit{K}$ is set to 7.}
\label{Fig6_K_ablation}
\end{figure}

During the graph construction stage of the ViG3D module, the number of vascular nodes determines the range of aggregation.
Insufficient neighbors can hinder effective information exchange, while too many can lead to over-smoothing \cite{han2022ViG}.
Therefore, selecting an appropriate number of neighbor nodes (denoted by $\mathit{K}$) is crucial for maintaining connectivity in the segmentation of 3D vascular structures.
As shown in Fig. \ref{Fig6_K_ablation}, adjusting $\mathit{K}$ from 3 to 32 resulted in the model achieving the highest DSC of 81.72 and the smallest HD95 of 15.21 when $\mathit{K}$ was set to 7.
This indicates that neither excessively high nor low values of $\mathit{K}$ enable optimal performance in the coronary segmentation task. At $\mathit{K}=7$ , the model achieves its highest spatial representation capability.

In addition to the number of neighboring nodes, the number of ViG3D blocks (denoted as $\mathit{L}$) is also crucial to capture detailed information about the vascular branch.
The parameter $\mathit{K}$ affects the efficiency of aggregation within a single-layer feature map, while the number of ViG3D blocks $\mathit{L}$ influences the connectivity among multi-layer feature maps. 
A larger $\mathit{L}$ indicates a wider range for searching connected vascular voxels.
To balance the scope and efficiency of 3D graph aggregation, the optimal number of stacked ViG3D block layers must be determined. The impact of different numbers of ViG3D blocks is examined by experimenting with models of varying sizes, from small to large. 
The segmentation performance of the network with different numbers of ViG3D blocks is shown in Table \ref{tab:vig_unit_Ablation}. The best DSC and HD95 performance is achieved when the number of ViG3D unit layers is configured to 2, 4, 16, and 2.

\begin{table}[!htbp]
\centering
\caption{Performance comparison with various ViG3D units on the ImagesCAS dataset.} 
\label{tab:vig_unit_Ablation} 
\resizebox{\columnwidth}{!}{
\begin{tabular}{lcccccc} 
\toprule 
\multirow{2}{*}{Model}&  \multicolumn{4}{c}{number of ViG3D units} &  \multirow{2}{*}{DSC ↑}& \multirow{2}{*}{HD95 ↓}  \\
                          \cline{2-5}
                      &   Enc\_3& Enc\_4 &Enc\_5 &Enc\_6                                                \\
 \midrule
ViG3DUNet-Small      &        1&       1&      4&      2&      81.54&                 15.83           \\ 
ViG3DUNet-Base       &        2&       2&      6&      2&      81.66&                 16.11           \\        
ViG3DUNet-Medium     &        2&       4&      6&      2&      81.53&                 17.02           \\
ViG3DUNet-Large      &        2&       4&      16&     2&      \textbf{81.72}&       \textbf{15.21}   \\
ViG3DUNet-Huge       &        2&       4&      32&     2&      76.94&                 40.95           \\
\bottomrule
\end{tabular}
}
\end{table}

The effectiveness of the structural modules in ViG3D-UNet was assessed through a series of structural ablation studies. The DSC and HD95 values are presented in Table \ref{tab:vig_unit_Ablation}.
Initially, the segmentation performance of models that exclude the ViG3D module was evaluated. Without the ViG3D module, the model relied solely on 3D convolution operations, effectively degrading to nnUNet \cite{isensee2021nnUnet}. Compared to the network incorporating the ViG3D module, there was a decrease of 0.61 in the DSC value and an increase of 2.39 in the HD95 value. Compared to STUNet \cite{huang2023Stu-net}, the overall performance of the model decreased without the ViG3D module.

Subsequent ablation studies were conducted on the Channel Attention (CA) module within the parallelized encoder. In the absence of channel attention, the features of CNN and ViG modalities were directly concatenated by the encoder. As indicated in the second row of Table \ref{tab:vig_unit_Ablation}, there was a decrease of 0.51 in the DSC value and an increase of 1.88 in the HD95 value.
This resulted in performance that was not competitive with networks that use the channel attention module, highlighting its importance for improving ViG3D-UNet performance.

\begin{table}[!htbp]
\centering
\caption{Performance comparison of structural ablation on the ImageCAS dataset.} 
\label{tab:vig_unit_Ablation} 
\resizebox{\columnwidth}{!}{
\begin{tabular}{lcccll} 
\toprule 
Method&      ViG3D&                CA&           OD&        DSC ↑&  HD95 ↓ \\
\midrule
\textbf{ViG3D-UNet}& \CheckedBox& \CheckedBox&  \CheckedBox& \textbf{81.72}&  \textbf{15.21}\\
\hline
-w/o ViG3D&   \XBox&       \XBox&  \XBox&         81.11 (↓0.61)&   17.60 (↑2.39)\\
-w/o CA &     \CheckedBox& \XBox&  \CheckedBox&   81.21 (↓0.51)&   17.09 (↑1.88)\\
-w/o OD &     \CheckedBox& \CheckedBox& \XBox&    80.56 (↓1.16)&   18.34 (↑3.13)\\
\hline
nnUNet\cite{isensee2021nnUnet}&       -&           -&               -&    81.11&   17.06\\
STUNet\cite{huang2023Stu-net}&       -&           -&               -&    81.54&   17.97\\
\bottomrule
\end{tabular}
}
\end{table}

Finally, the segmentation performance without the Offset Decoder (OD) mechanism was analyzed. As shown in Table \ref{tab:vig_unit_Ablation}, the ablation model demonstrates poorer performance compared to the proposed method, with a decrease of 1.16 in the DSC value and an increase of 3.13 in the HD95 value.
A noticeable decline in performance was also observed in the model without the offset decoder compared to other approaches \cite{isensee2021nnUnet,huang2023Stu-net}. 
This indicates that denser skip connections in ViG3D-UNet hinder improvements in segmentation performance. The ablation experiments confirm the significance of each module within the network.

\subsection{Performance Comparison}
Experiments were performed on the ASOCA and ImageCAS datasets to evaluate ViG3D-UNet against four leading segmentation models. 
The nnU-Net \cite{isensee2021nnUnet} is a self-configuring approach for deep learning-based medical image segmentation, noted for its exceptional performance across diverse tasks without needing manual adjustment. 
STUNet \cite{huang2023Stu-net}, chosen for its comparable model weight scale to ViG3D-UNet, is a transfer learning network featuring scalable and adaptable U-Net parameters. 
UNETR \cite{hatamizadeh2022unetr} utilizes a hybrid architecture combining a transformer encoder with a convolutional decoder.
SwinUNETR \cite{hatamizadeh2021SwinUnetr} employs the Swin Transformer \cite{liu2021swin-t} as encoders for feature extraction, establishing itself as a prominent baseline in medical transformer segmentation. 
To ensure a fair assessment of each model's capabilities, the use of post-processing was minimized across all methods.

\begin{table}[!htbp]
\centering
\caption{Performance comparisons on ASOCA and ImageCAS dataset with the state-of-the-art algorithms.  
)} 
\label{tab:metrics} 
\resizebox{\columnwidth}{!}{
\scalebox{1}[1]{
\begin{tabular}{clccccc} 
\toprule 
 Dataset&  Method             & DSC ↑ & HD95 ↓  & IoU ↑  & Prec ↑ & Recall ↓\\
\midrule 
\multirow{5}{*}{ASOCA}&nnUNet\cite{isensee2021nnUnet} & 80.53 & 35.93     & 65.04  & 81.05  &80.99  \\

           &STUNet\cite{huang2023Stu-net}            & \textbf{84.80}  & 16.83     & \textbf{73.81}  & 88.36  & \textbf{82.31} \\

           &UNETR\cite{hatamizadeh2022unetr}             & 76.95  & 35.60    & 63.51  & 75.76  &78.34  \\

           &SwinUNETR\cite{hatamizadeh2021SwinUnetr}         & 79.96  & 27.99    & 66.78  & 80.01  &  81.12\\

           &ViG3D-UNet             & 84.23  & \textbf{13.39}     &73.06  & \textbf{90.39}  & 79.65 \\
\hline
\multirow{5}{*}{ImageCAS}&nnUNet\cite{isensee2021nnUnet} & 81.11 & 17.60 & 69.31   & 81.70  & 82.04 \\
            &STUNet\cite{huang2023Stu-net}            & 81.54  & 17.97     & 69.17  & 80.21  & 83.51 \\

            &UNETR\cite{hatamizadeh2022unetr}             & 76.57  & 28.53    & 62.39  & 72.03  & 82.33 \\

            &SwinUNETR\cite{hatamizadeh2021SwinUnetr}         & 76.10  & 41.64    & 61.08  & 70.58  & \textbf{83.22} \\

            &ViG3D-UNet             & \textbf{81.72}  & \textbf{15.21}    & \textbf{69.40}  & \textbf{83.16}  & 80.79 \\
\bottomrule 
\end{tabular}
}   
}   
\end{table}

\begin{figure}[!htbp]
\begin{center}
   \includegraphics[width=0.9\linewidth]{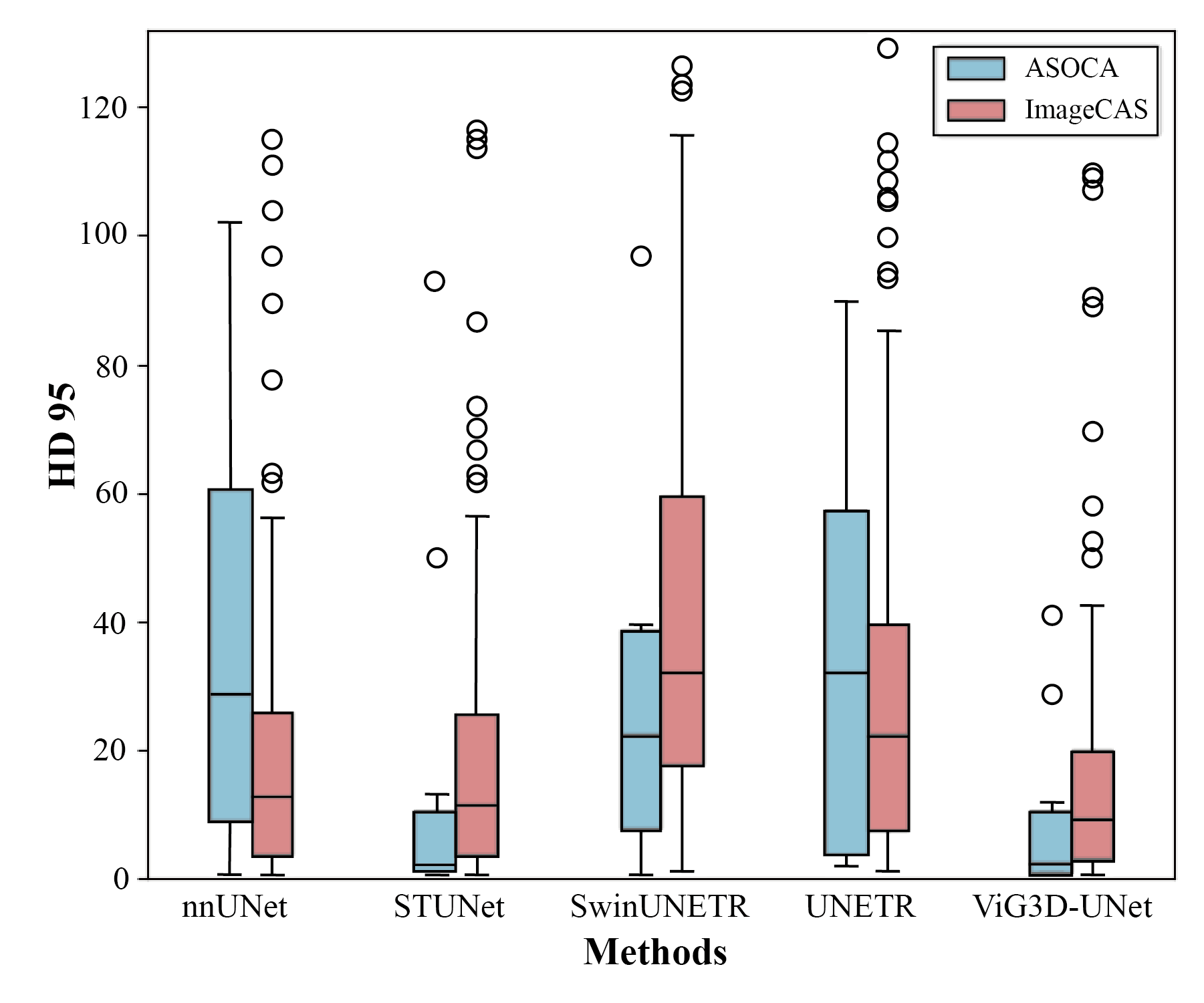}
\end{center}
   \caption{Box plot of HD95 values for individual cases across different methods. Outliers, or data points that deviate significantly from the HD95 mean, are marked with circles}
\label{fig5_hd95_box}
\end{figure}

\begin{figure*}[!htbp]
\centering
\includegraphics[width=0.95\linewidth]
{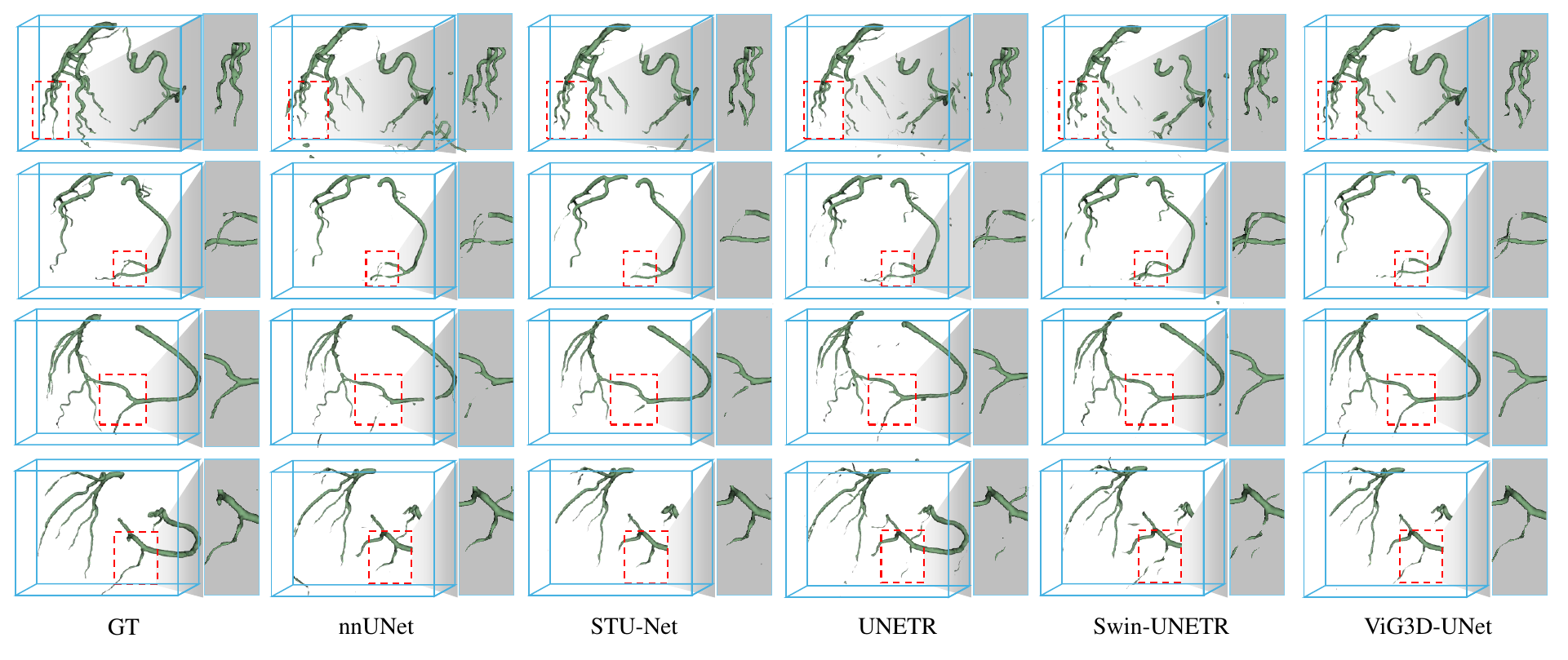}
\caption{In the visual comparison on the small-scale ASOCA dataset, our method yields more complete segmentation results at the terminal branches of the coronary arteries. }
\label{fig3_ASOCA_results}
\end{figure*}

\begin{figure*}[!htbp]
\centering
\includegraphics[width=0.95\linewidth]
{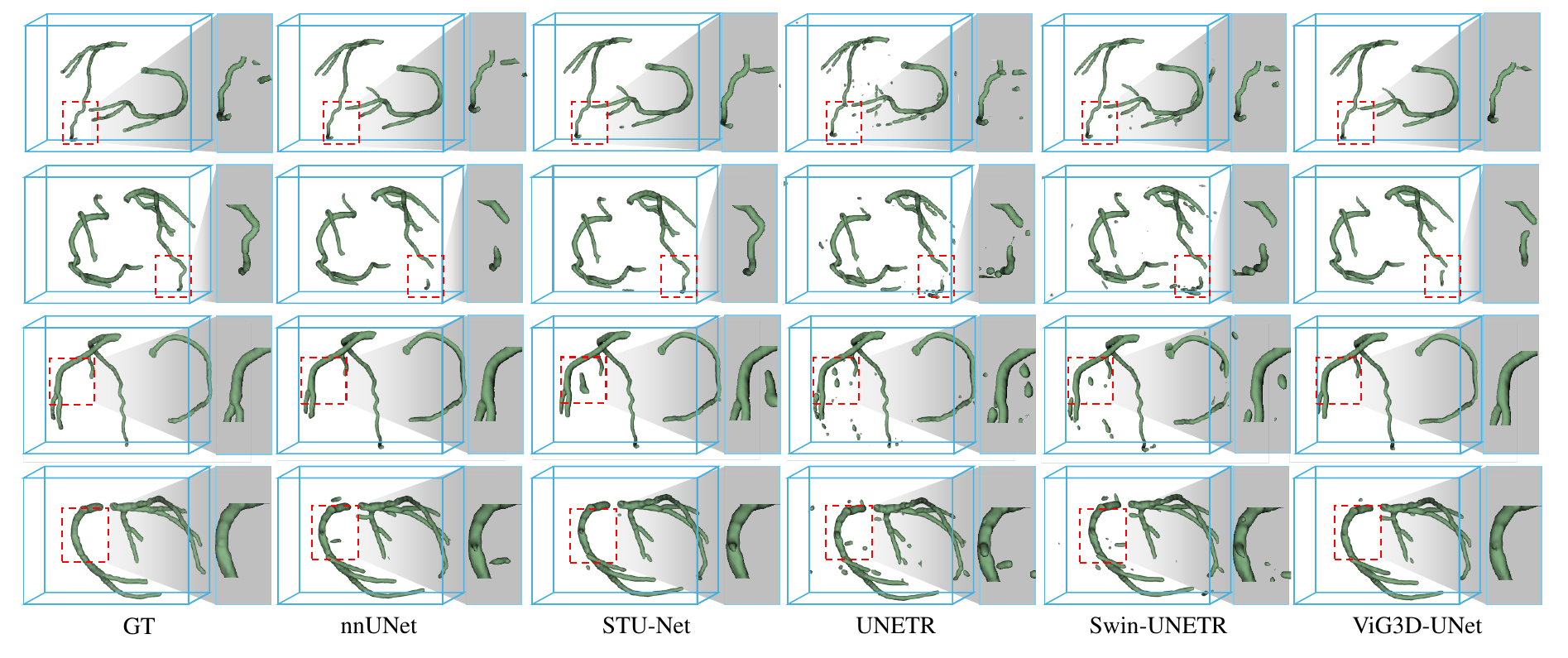}
\caption{In the visual comparison on the ImageCAS dataset, our method maintains greater connectivity at the coronary artery endpoints.  Even when segmentation breaks occur, it results in fewer disconnected voxels compared to other methods. Additionally, the segmentation results demonstrate fewer mis-segmentations compared to other algorithms.}
\label{fig4_ImageCAS_results}
\end{figure*}

Quantitative comparisons with state-of-the-art methods are shown in Table \ref{tab:metrics}. 
In the ASOCA dataset, the proposed method achieved a DSC of 84.23\%, a HD95 of 13.39, an IoU of 73.06\%, a precision of 90.39\%, and a recall of 79.65\%. The HD95 and precision metrics are the best among the methods compared, with the HD95 being 3.44 lower and the precision 2.03 higher than the second-best model.
The HD95 metrics demonstrate that the proposed method achieves the smallest boundary discrepancy with the ground truth, indicating superior connectivity in the segmentation outcomes. 
The precision indicates that the proposed method has fewer mis-segmentations compared to other methods.
The DSC of ViG3D-UNet is second, just 0.57 behind STUNet \cite{huang2023Stu-net}, while significantly outperforming other CNN-based \cite{isensee2021nnUnet} and transformer-based \cite{hatamizadeh2022unetr, hatamizadeh2021SwinUnetr} approaches. 

In the large-scale ImageCAS dataset, the evaluation results in Table \ref{tab:metrics} show that the proposed method achieves a dice similarity coefficient (DSC) of 81.72\%, an HD95 of 15.21, an IoU of 69.40\%, precision of 83.16\%, and recall of 80.79\%.  
The DSC of ViG3D-UNet exceeds that of current state-of-the-art methods by 0.57, highlighting its superior accuracy in segmentation. 
Furthermore, the method shows enhanced HD95 performance, with a value 2.39 lower than the second-best DSC score obtained by STUNet \cite{huang2023Stu-net}.
This indicates its capability to maintain closer alignment with ground truth boundaries, while also improving segmentation connectivity.
The proposed method achieves the best values in DSC, HD95, IoU, precision and recall. Consequently, this analysis confirms that ViG3D-UNet delivers state-of-the-art performance on the ImageCAS dataset when compared to existing methods.

To further investigate the differences in boundary error and connectivity among various methods, we performed a statistical analysis of the HD95 metric on two separate datasets. 
The HD95 distribution of individual cases across different methods in the ASOCA dataset is analyzed, as shown by the blue boxes in Fig. \ref{fig5_hd95_box}. 
The ViG3D-UNet exhibits a more compact HD95 distribution, with outliers closer to the mean value, suggesting enhanced segmentation connectivity and model robustness.
In the ImageCAS dataset, as shown by the red boxes in Fig. \ref{fig5_hd95_box}, our method also exhibits a similar statistical distribution of HD95 compared to other methods. 
Our approach, shown on the far right, has a distribution closer to the mean with fewer outliers compared to the other methods. 
This pattern in the HD95 distribution indicates improved segmentation connectivity and robustness, which confirms enhanced segmentation connectivity and accuracy.

\section{Discussion} 
CTA is widely used for the screening of cardiovascular diseases due to its non-invasive nature, lower risk, and lower cost.
Coronary segmentation, as a crucial step in vascular reconstruction, is essential for the diagnosis of cardiovascular diseases and the planning of surgeries.
Current medical image segmentation methods often fall short in producing continuous vascular structures. 
To address this issues, an end-to-end graph-based U-shaped volumetric segmentation framework is proposed to enhance vessel connectivity while achieving high accuracy.

Vascular morphology varies between individuals, but human coronary vessels exhibit a consistent topological structure.
To enhance model connectivity, a 3D vision GNN module is designed to represent and aggregate features of vascular topological structures (Fig. \ref{fig2_3DVIG}). 
A dual-branch parallelized encoder and a paperclip-shaped offset decoder are introduced for feature integration and accurate segmentation.
The two main contributions of ViG3D-UNet are as follows: (1) The 3D vision GNN module, ViG3D, is tailored to capture and aggregate spatial graph features, with vascular graphs automatically constructed during training to extract topological structures within an end-to-end segmentation framework. The results in Tables \ref{tab:metrics_ours} and \ref{tab:vig_unit_Ablation} demonstrate its effectiveness.
(2) The paperclip-shaped offset decoder integrates textural features of CNNs with topological features from vision GNNs, addressing the challenge of direct concatenation, which often leads to high computational costs and suboptimal feature fusion. As shown in Table \ref{tab:vig_unit_Ablation}, model performance is significantly improved by this module.

The ViG3D-UNet model has been shown to have significant connectivity advantages over other state-of-the-art methods, as indicated in Table \ref{tab:metrics} and Fig. \ref{fig5_hd95_box}.
Individual case visualizations are presented in Fig. \ref{fig3_ASOCA_results} and Fig. \ref{fig4_ImageCAS_results}. 
For example, the second row of Fig. \ref{fig3_ASOCA_results} reveals that while all methods exhibit vascular breaks, only one break with intact branches is shown by ViG3D-UNet, in contrast to others that display multiple breaks or incomplete major branch segmentation.
Fig. \ref{fig4_ImageCAS_results} further demonstrates that ViG3D-UNet achieves more complete vascular segmentation, with fewer miss-segmented voxels and a more continuous segmentation mask at vascular endpoints.
Additionally, fewer breaks and mis-segmentations are consistently shown by ViG3D-UNet. 
In both datasets, vessel structures are effectively leveraged, enhancing connectivity in vascular branches.

CTA is often performed as an outpatient procedure and requires fewer recovery times, which makes it more convenient for patients. However, its precision in assessing vascular morphology is limited compared to coronary digital subtraction angiography \cite{manubolu2024innovations,2018EuropeanHeartJournal}. This limitation is evident in the discrepancies between the diameters of the reconstructed vessels and their actual anatomical structures, as well as in the fragmentation of vessels that are expected to appear continuous. The accuracy of the diagnosis of cardiovascular disease using CTA can be further enhanced by the proposed method. This improvement enables the more frequent use of cost-effective and lower-risk CTA as a reliable imaging tool evaluate coronary artery disease in outpatient and screening settings.

Despite  promising results, several limitations are still present in the proposed method. First, the model is trained from scratch in scenarios with limited sample sizes. Improved segmentation accuracy could be achieved by pre-training the backbone on a large dataset. Second, a blind search of feature distances is involved in the current spatial graph representation. Given that blood vessels are irregular fractal structures \cite{Masters2004Fractal}, the incorporation of structural knowledge could enhance the efficiency of topological feature extraction.

\section{Conclusion}
In this study, ViG3D-UNet is introduced as a 3D vision graph neural network framework for the continuous segmentation of the coronary arteries. 
By incorporating the ViG3D module, the traditional planar Vision GNN is adapted to process and aggregate volumetric image features. 
The ViG3D-UNet framework consists of a dual-branch parallelized encoder and a paperclip-shaped offset decoder, designed to effectively leverage the topological morphology of vascular structures. 
ViG3D-UNet demonstrates superior accuracy and connectivity in coronary segmentation, outperforming existing state-of-the-art approaches. 
In general, this study significantly contributes to the development of more accurate and seamlessly integrated medical image segmentation frameworks.

\bibliographystyle{IEEEtran}
\bibliography{ref}

\end{document}